\documentclass[prl,aps,amsfonts,twoside,twocolumn,amssymb,superscriptaddress,reprint]{revtex4}

\usepackage[latin1]{inputenc}
\usepackage[english]{babel}
\usepackage[draft]{hyperref}
\usepackage{graphicx}
\usepackage{amsmath}
\usepackage{units}
\usepackage{color}
\usepackage{float}
\usepackage[caption=false]{subfig} 
\usepackage{blindtext}
\begin{document}

\title{Continuous-wave squeezed states of light via `up-down' self-phase modulation}

\author{Amrit Pal Singh}
\affiliation{Institut f\"ur Laserphysik \& Zentrum f\"ur Optische Quantentechnologien,  Luruper Chaussee $149$, $22761$ Hamburg, Germany}
\author{Stefan Ast}
\author{Moritz Mehmet}
\author{Henning Vahlbruch}
\affiliation{Max Planck Institute for Gravitational Physics (Albert Einstein Institute) and Leibniz Universität Hannover, Callinstra{\ss}e $38$, $30167$ Hannover, Germany}
\author{Roman Schnabel}
\affiliation{Institut f\"ur Laserphysik \& Zentrum f\"ur Optische Quantentechnologien,  Luruper Chaussee $149$, $22761$ Hamburg, Germany}

\begin{abstract} 
Continuous-wave (cw) squeezed states of light have applications in sensing, metrology and secure communication. In recent decades their efficient generation has been based on parametric down-conversion, which requires pumping by externally generated pump light of twice the optical frequency. Currently, there is immense effort in miniaturizing squeezed-light sources for chip-integration. Designs that require just a single input wavelength are favored since they offer an easier realization. Here we report the first observation of cw squeezed states generated by self-phase modulation caused by subsequent up and down conversions. The wavelengths of input light and of balanced homodyne detection are identical, and 1550\,nm in our case. At sideband frequencies around $1.075\,$GHz, a nonclassical noise reduction of $(2.4\pm 0.1)\,$dB is observed. The setup uses a second-order nonlinear crystal, but no externally generated light of twice the frequency. Our experiment is not miniaturized, but might open a route towards simplified chip-integrated realizations.
\end{abstract}

\maketitle

\begin{figure}[]
    \includegraphics[width=8.7cm]{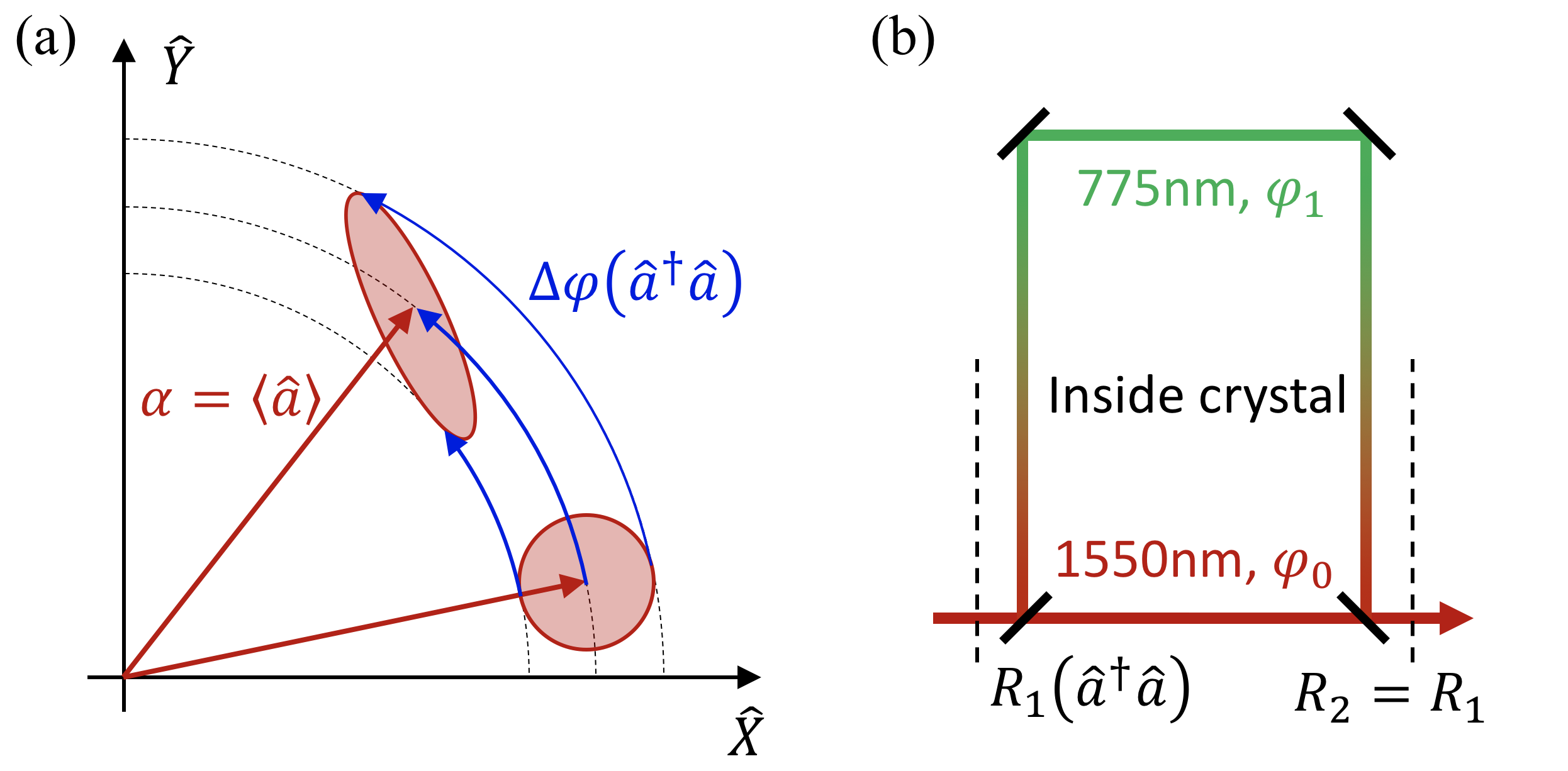}
    \caption{(a) Phase-space description of the transformation from a coherent state to a squeezed state through self-phase modulation. $\hat X$ and $\hat Y$ are the amplitude and phase quadrature amplitude operators defined with respect to a local oscillator field. The shaded areas correspond to quantum uncertainties. The phase shift depends on the light's intensity, which is proportional to the photon number operator $\hat n = \hat a^\dagger \hat a$. The latter are the creation and annihilation operators. (b) Illustration of how the intensity dependent phase shift is realized in our setup. During transmission through a PPKTP crystal some fraction of the light, here represented by the reflectivity $R_1$ of a fictitious mirror, is frequency up-converted and subsequently re-converted. In this case, the optical propagation phases $\varphi_1$ and $\varphi_0$ differ. The converted fraction $R_1$ depends on $\hat n$.}
    \label{fig:1}
\end{figure}

\paragraph{Introduction}-- 
Squeezed states of continuous-wave (cw) quasi-monochromatic light have been improving the sensitivity of the gravitational-wave detector GEO\,600 during its observational runs since 2010 \cite{LSC2011,Grote2013} and were tested for integration in LIGO \cite{LSC2013}. More recently, proof-of-principle experiments used the unique feature of these states for one-sided device independent quantum key distribution (QKD) \cite{Gehring2015}, oblivious transfer (OT) \cite{Furrer2018}, and the calibration of photo-diode quantum efficiencies without knowledge of the incident light power \cite{Vahlbruch2016}. Noteworthy, the latter three experiments are impossible by scaling the optical power of light in coherent states.\\
State-of-the-art devices for the generation of squeezed states of light are based on cavity-enhanced (type\,I) degenerate parametric down-conversion (PDC) 
in a nonlinear crystal such as MgO-doped LiNbO$_3$ or quasi-phase-matched periodically poled KTP \cite{Wu1986,Vahlbruch2008,Schnabel2017}. Here, the nonlinear process is of second order and requires pump light at twice the optical frequency, which needs to be spatially overlapped with the cavity mode of fundamental frequency. The fine-tuning of phase-matching between the two wavelengths is realised via the temperature of the crystal.  
The largest squeeze factors observed so far are $15$\,dB at $1064$\,nm and $13$\,dB at 1550\,nm below vacuum noise  \cite{Vahlbruch2016,Schonbeck2018}. More than $10$\,dB of two-mode-squeezing, i.e.~the continuous-variable entanglement between two beams of light was realised in \cite{Eberle2013}. The limitations to the squeeze factors were in all cases set by photon loss.  
The states were produced in about 10\,mm long crystals surrounded by $4$\,cm long standing-wave 
cavities, and coupled to freely propagating beams. The complete optical setups had footprints of one to two square meters to allow for individually optimized mode-matchings and electro-optical phase controls \cite{Vahlbruch2010}. Commercial applications of squeezed light, however, demand more compact devices. Micro-optomechanical devices and integrated optics could offer ways for miniaturization \cite{Safavi-Naeini2013,Purdy2013b,Dutt2015,Hoff2015,Gehring2017}.\\  
Other approaches for the generation of squeezed light are non-degenerate four-wave mixing \cite{Slusher1985,Shelby1986} and self-phase modulation (fully degenerate four-wave mixing). 
Self-phase modulation (SPM) has the convenient feature that just a single optical frequency needs to be supplied for the generation and observation/exploitation of squeezing. Fig.\,\ref{fig:1}\,(a) illustrates the self-transformation from a coherent state to a squeezed state via SPM when propagating through a medium having an intensity-dependent refractive index. The latter corresponds to the optical Kerr effect and was theoretically analysed in \cite{Yurke1984,Reynaud1989,Thuring2011}. 
SPM is based on a third-order nonlinear coupling of light with itself, which requires high intensity. Ref.~\cite{Bergman1991} reported $5$\,dB of squeezing due to SPM on light pulses after propagation through a $50$\,m glass fibre loop. Ref.~\cite{Dong2008} observed $6.8$\,dB on light pulses after propagation through a $13$\,m fibre. The generation of squeezed states in the cw regime is more challenging. 
SPM due to radiation pressure on a micro-mechanical device was recently used to produce cw squeezed light. About $0.2$\,dB and $1.7$\,dB were achieved, respectively \cite{Safavi-Naeini2013,Purdy2013b}. 
These optomechanical experiments required operation in vacuum and cryogenic cooling.\\ 
So far cw squeezing from all-optical SPM was only achieved in the cascaded second-order nonlinear \cite{Ostrovskii1967,DeSalvo1992,White1996a} `down-up' setting \cite{Kasai1997,Zhang2001}. These experiments realized SPM via cavity-enhanced above-threshold \emph{down}-conversion with subsequent sum-frequency \emph{up}-conversion, i.e.~`down-up' SPM. About $1.5$\,dB of squeezing 
was generated, respectively. This value, however, was not observed with respect to the shot noise level of the detector's local oscillator beam but only with respect to a slightly higher shot noise level that was calculated and that included the contribution from the non-negligible signal beam power. A general challenge with `down-up' SPM is the required above-oscillation-threshold operation \cite{Fabre1990}, and a drawback that on cavity resonance just up to $3$\,dB can be produced, even when assuming zero optical loss and an intra-cavity intensity as high as four-times the oscillation threshold \cite{Fabre1990}.  
The direct observation of cw squeezed states of light generated by all-optical SPM below parametric oscillation threshold has not been achieved so far. A suitable approach is given by `up-down' SPM \cite{Kasai1997}, which was previously used to reduce classical fluctuations of laser light \cite{White2000,Khalaidovski2009}. The effect per cavity round-trip corresponds to that of degenerate PDC and in principle an arbitrarily high squeeze factor can be produced in the loss-less case without surpassing the oscillation threshold \cite{Yurke1984}. 

Here, we report the direct observation of squeezed states of light at the telecommunication wavelength of $1550$\,nm produced by all-optical self-phase modulation at intensities below the oscillation threshold using `up-down' SPM. A $70$\,mW beam that initially was in a coherent state transformed itself to a $(2.4\pm 0.1)\,$dB squeezed state by transmission through a $10$\,mm long periodically poled KTP (PPKTP) crystal inside a traveling-wave cavity. The temperature of the crystal was precisely adjusted (around the design temperature of about $60^\circ$C) to a \emph{minimum} of the sinc-squared function to prevent second-harmonic generation. A second cavity subsequently filtered the undepleted $70$\,mW light. The remaining field had a negligible power and was observed with conventional balanced homodyne detection.

\paragraph{Experimental implementation}--
Fig.\,\ref{fig:1}\,(b) illustrates the fundamental principle of `up-down' self-phase modulation. First of all, the PPKTP-crystal temperature is set to a conversion minimum next to the global conversion maximum. At such a temperature (almost) no up-converted light leaves the crystal. Light that is up-converted from $1550$\,nm to $775$\,nm during propagation through the first half of the crystal is fully converted back to $1550$\,nm during propagation through the second half of the crystal. The conversion efficiency at the crystal's centre depends on the intensity of the $1550$\,nm field, which is a variable of non-zero quantum uncertainty. The unmatched propagation phases together with the intensity-dependent conversion efficiency generate an intensity dependent phase shift of the 1550\,nm light beam, which results in self-phase modulation. 
\begin{figure}[]
    \includegraphics[width=8.5cm]{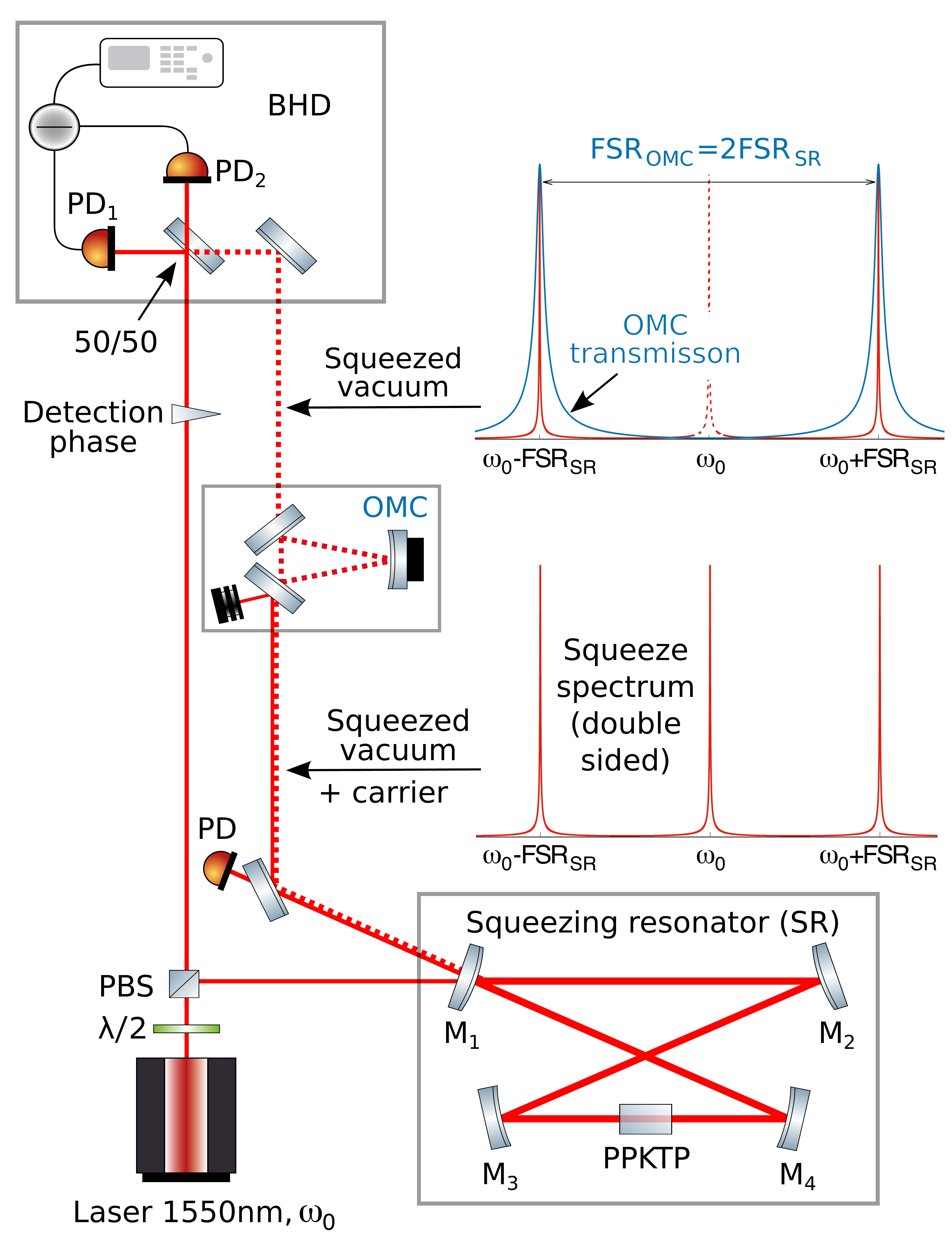}
    \caption{Schematic of the experimental setup. The squeezing resonator had bow-tie geometry and was servo-controlled on resonance for the input light. Squeezed states were generated via all-optical `up-down' SPM in PPKTP. The carrier beam (solid red line) and squeezing given by entangled  pairs at frequencies of plus/minus an odd multiple of the squeezing resonator's FSR (dashed red line) were separated by an output mode-cleaner (OMC).  Lower spectrum: low-power built-up in the squeezing resonator versus its detuning. Upper spectrum: Filter effect of the OMC resonator. PZT: piezoelectric transducer, $\lambda/2$: half-wave plate, PBS: polarizing beam splitter, PD: photo diode. $\omega_0$: optical angular frequency of the quasi-monochromatic light.}
    \label{fig:2}
\end{figure}

Fig.\,\ref{fig:2} shows the schematic of our setup. The main laser was a $5$\,W erbium-doped continuous-wave single-frequency $1550\,$nm fiber seed laser (model \textit{Koheras BOOSTIK}). Our experiment required about $100$\,mW. Most of it was mode-matched to the squeezing resonator, and $4$\,mW served as local oscillator for balanced homodyne detection (BHD). The squeezing resonator was a bow-tie resonator with free spectral range (FSR) of about $358\,$MHz, i.e.~with round trip length of about $83.8\,$cm. It contained a PPKTP crystal with dimensions $9.3\times 2.0\times 1.0\,$mm$^{3}$. The small end faces had anti-reflection coatings. The coupling mirror $\text{M}_{1}$ had a power reflectivity of $99\,\%$ at $1550\,$nm, while the other mirrors were highly reflective. 
All mirrors had a power reflectivity of $20\,\%$ at $775\,$nm. The radius of curvature of the convex mirrors $\text{M}_{1}$ and $\text{M}_{2}$ was $-500\,$mm. Together with the concave mirrors $\text{M}_{3}$ and $\text{M}_{4}$, which had radii of curvatures of $100\,$mm, a waist of $30\,\mu$m was formed in the centre of the PPKTP crystal. The produced squeezed sidebands and the bright carrier field left the bow-tie resonator in reflection and were spatially separated by an output mode-cleaner (OMC). The FSR of the OMC was precisely twice that of the squeezing resonator and was fine adjusted manually with a high-voltage driven piezo-mounted mirror. The filtering by the OMC had the purpose to produce a signal field without carrier to allow for (conventional) BHD. Our filtering scheme prevented the detection of frequencies within the linewidth of the squeezing resonator and at frequencies of even number multiples of the squeezing resonator's FSR. Squeezed vacuum states at frequencies of odd multiples of the squeezing resonator's FSR  propagated to the BHD where they were superimposed with a local oscillator at a $50/50$ beam splitter. The BHD was home-built \cite{AstS2016} and had a bandwidth of $1.5\,$GHz allowing for the detection of squeezed states at about $358\,$MHz and $1074\,$MHz.

\begin{figure}[]
    \includegraphics[width=8.8cm]{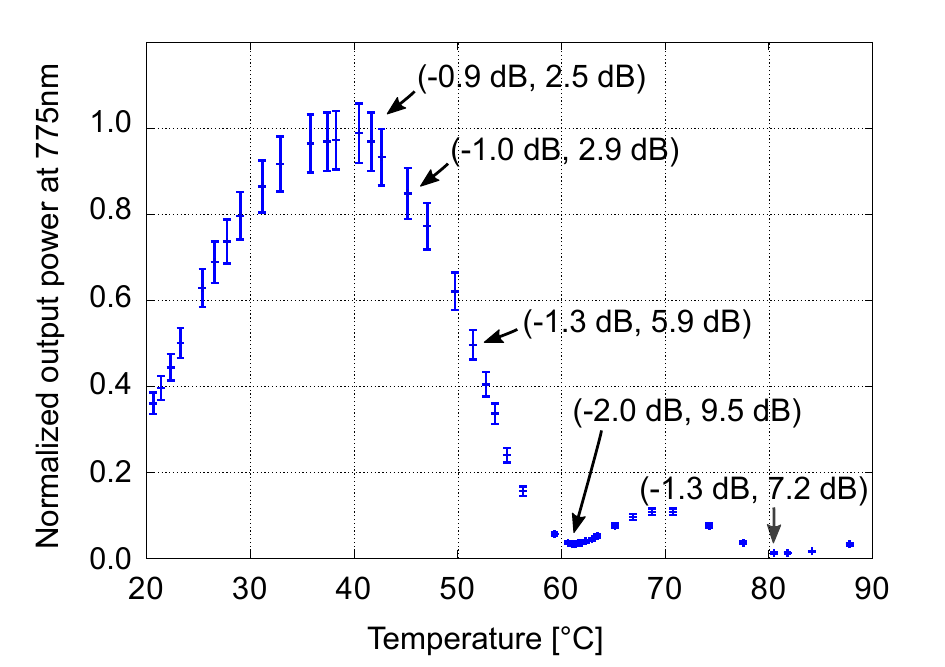}
    \caption{Observed squeeze factors (negative sign) and anti-squeeze factors observed at the sideband frequency of $358$\,MHz. The highest squeezed factor was achieved in the first total (up-down) conversion minimum and is attributed to self-phase modulation. Also at the conversion maximum as well as at other operational points of the experiment squeezed states were produced. This is due to another process, namely \emph{the nonlinear depletion} of the fundamental wave, also called \emph{SHG squeezing}, which was previously demonstrated in Refs.~\cite{Pereira1988,Kurz1993}.}
    \label{fig:5}
\end{figure}

\paragraph{Measurement results}--
\begin{figure}[]
    \includegraphics[width=8.5cm]{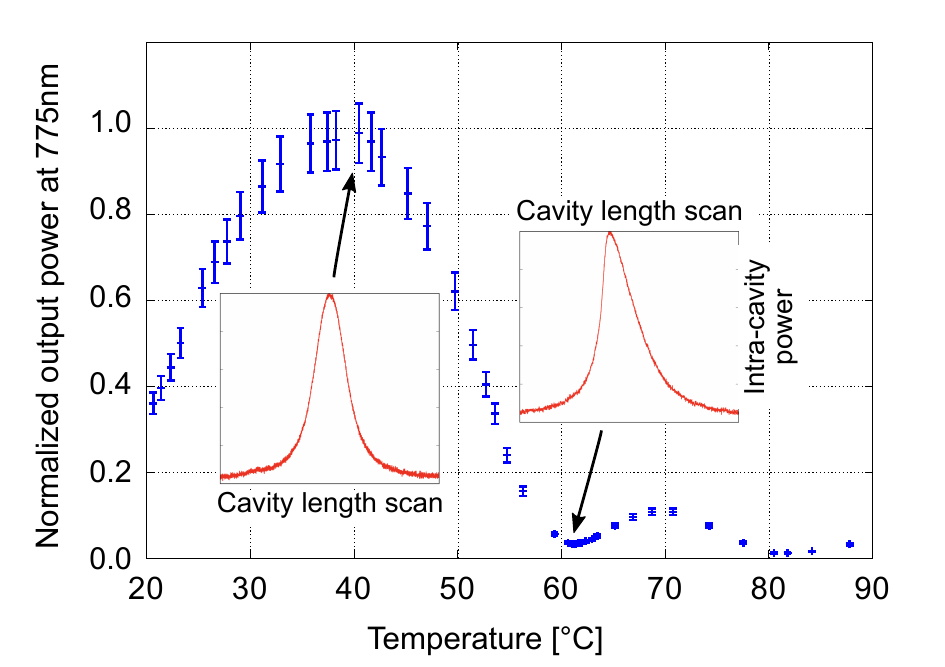}
    \caption{Measurement of the second-harmonic conversion efficiency as a function of crystal temperature. Efficient `up-down' SPM is attributed to conversion minima, here at $61.2\,^{\circ}$C and $81.8\,^{\circ}$C. The global (up-) conversion maximum was found at $40.5\,^{\circ}$C. The insets show the measured resonance profiles for linear scans of the cavity length. In the conversion maximum the resonance profile was symmetric. At other temperatures, the phases of 1550\,nm and 775\,nm light were less-well matched and SPM resulted in asymmetric peaks, with maximal asymmetry in the conversion minima. As predicted by theory, the asymmetries were found to be more pronounced for higher light intensities and were independent of scanning speeds.}
    \label{fig:3}
\end{figure}
Self-phase modulation with lowest loss is achieved at crystal temperatures at which the SHG efficiency is minimal.
To identify these, we varied the temperature of the crystal between $20\,^{\circ}$C and $88\,^{\circ}$C while having an input light power of $8.8\,$mW at $1550\,$nm, see Fig.~\ref{fig:3}. For all measurement points, the length of the squeezing resonator was stabilized on resonance with the Pound-Drever-Hall (PDH) method. Light converted to $775\,$nm was coupled out via mirror $\text{M}_{3}$, passed through a dichroic beam splitter to filter $1550\,$nm light, and was detected with a power meter. The global SHG conversion maximum was found at $40.5\,^{\circ}$C. The first and the second high-temperature-sided conversion minima were observed at $61.2\,^{\circ}$C and $81.8\,^{\circ}$C, respectively. 
At conversion minima linear scans of the cavity length revealed significantly deformed resonance profiles, as shown in Fig.~\ref{fig:3} (right inset). At $40.5\,^{\circ}$C the resonance profile was not deformed (left inset). For these measurements we recorded the out-coupled power at 1550\,nm behind mirror $\text{M}_{2}$. 
The shape of the resonance profiles did not depend on the scanning speed (which was of the order of microseconds per full width at half maximum). This result showed that absorbed light in combination with the crystal's thermal expansion coefficient did not have an influence on the resonance profiles. 
\begin{figure}[]
    \includegraphics[width=7.7cm]{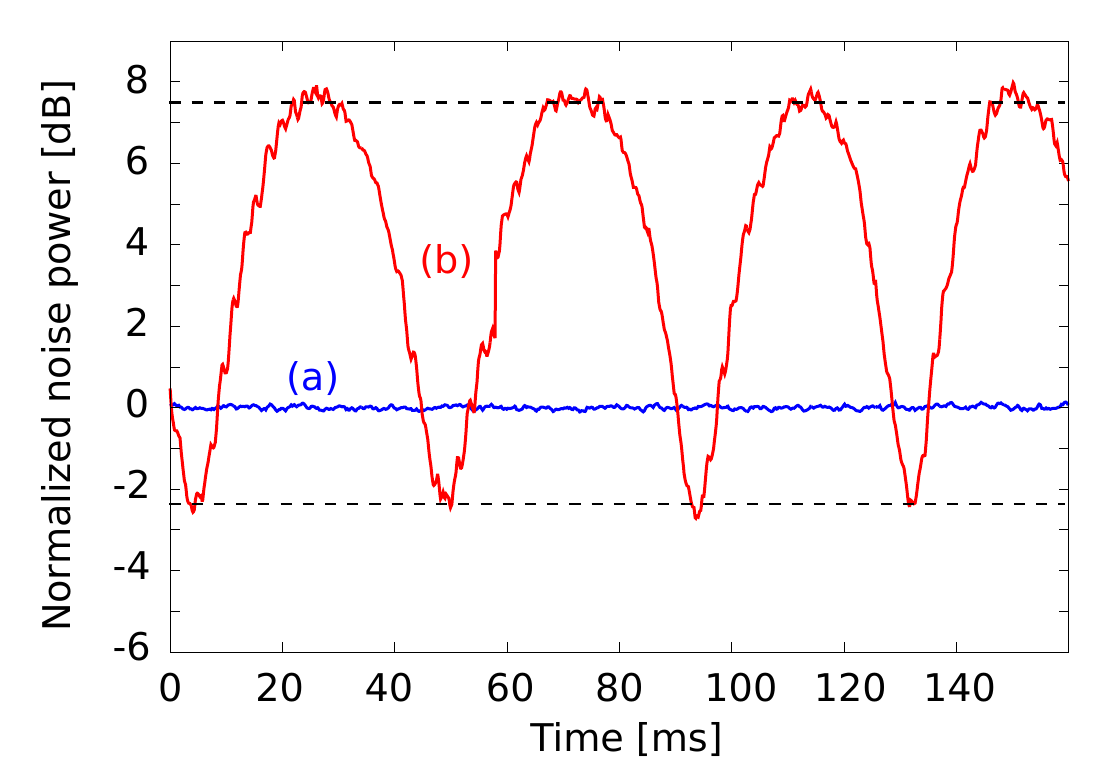}
    \caption{Tomographic measurement on a squeezed state produced by SPM. Trace (a) serves as the reference and represents the vacuum noise variance for a local oscillator power of 4\,mW. Trace (b) shows the zero-span noise variance of the squeezed state at a sideband frequency of $1074\,$MHz when the quadrature phase angle is scanned. At $1074\,$MHz we observed the highest squeezed factor in our setup of $\left(2.4\pm 0.1\right)\,$dB. The resolution bandwidth was set to $500\,$kHz and the video bandwidth to $200\,$Hz. The electronic dark noise was at $-8.2\,$dB (not shown) and was not subtracted from the traces.}
    \label{fig:4}
\end{figure}
For squeezing generation, the temperature of the PPKTP crystal was adjusted to $61.2\,^{\circ}$C, and $70\,$mW of $1550$\,nm light was mode-matched to the squeezing resonator, whose length was stabilized close to resonance. The bright light field and the squeezed sidebands left the bow-tie resonator at mirror $\text{M}_{1}$ and were separated by the OMC. The remaining squeezed vacuum field was overlapped with the local oscillator beam on the $50/50$ beam splitter were a fringe visibility of $97\,\%$ was achieved. 
Fig.\,\ref{fig:4} represents the highest squeeze factor we reproducibly observed. Trace (a) is the measured vacuum noise level of our local oscillator power of $4\,$mW, which was recorded when the signal path of the BHD was blocked. With the signal path open the variance of the squeezed field was evaluated with a zero-span measurement at a sideband frequency of $1074\,$MHz, which corresponded to the third FSR of the bow-tie resonator. During the measurement, the relative phase between the signal field and the local oscillator was periodically changed with a piezo-actuated mirror that was located in the local oscillator path. 
As depicted by trace (b) in Fig.\,\ref{fig:4}, a nonclassical noise reduction of $(2.4\pm 0.1)\,$dB below the vacuum noise level and an anti-squeezing value of $(7.5\pm 0.1)\,$dB was detected. If optical loss was the only decoherence process, these values would request a total detection efficiency of just $47\,\%$ \cite{Schnabel2017}. Our independent measurements of the squeezing resonator's escape efficiency ($(84 \pm 2)\,\%$), the OMC transmission ($(89 \pm 1)\,\%$), remaining SHG ($(98 \pm 1)\,\%$), and the detection efficiency of BHD ($(90 \pm 4)\,\%$) suggested, however, a total detection efficiency of about $(66 \pm 5)\,\%$.\\
%
%
At the sideband frequency of the \emph{first} FSR, we observed (in the same conversion minimum) $(2.0\pm 0.1)\,$dB of squeezing together with $(9.5\pm 0.1)\,$dB of anti-squeezing, see Fig.~\ref{fig:5}. For this measurement we increased the input power to $85$\,mW. We could not observe any higher squeeze factor at $358$\,MHz. The difference of the two squeeze factors is not easy to explain since the optical loss and phase noise \cite{Franzen2006} were identical in these measurements. Other parameters such as the performance of the balanced homodyne detector at these frequencies were also identical.\\
A potential explanation for the higher squeeze factor at higher frequency is thermally-driven \emph{internal} phase noise \cite{Cesar2009}. Thermal energy results in microscopic vibration inside the (nonlinear) medium that translates into broadband random phase modulations of the carrier light, which usually falls off with frequency \cite{Callen1952,Harry2002a}. In one of the earliest fibre squeezing experiments cooling the fibre down to $4$\,K reduced noise and allowed the observation of squeezed states \cite{Shelby1986}. We had observed such noise in a previous similar setup \cite{Khalaidovski2009} and were able to describe it as broadband phase noise with a $1/f$ slope of its noise power \cite{Thuring2011}. The limitation of the squeeze factors in this work, however, cannot be described by the same slope and we could not clearly confirm the presence of internal phase noise.   

Fig.~\ref{fig:5} shows the highest achievable squeeze factors at $358$\,MHz for different temperatures of the PPKTP crystal. The highest squeeze and anti-squeeze factors are in the first conversion minimum. This observation clearly supports our claim that the dominant nonlinear process for squeezed state generation at this temperature was self-phase modulation. At other temperatures, where a significant amount of light power at $775$\,nm was coupled out, squeezed states were partly produced by non-linear depletion of the $1550$\,nm light. In the conversion maximum the latter effect was the only one. Non-linear depletion was previously used for the generation of squeezed states as reported in \cite{Pereira1988,Kurz1993}.

\paragraph{Summary and conclusion}--
We report the first generation of a continuous-wave squeezed field via self-phase modulation through subsequent second-order up and down conversion below oscillation threshold. The carrier light of $70\,$mW at $1550\,$nm was subsequently subtracted by a filter cavity. We directly observed a nonclassical noise-reduction of up to 2.4\,dB. Since the squeezed field was not accompanied by carrier light of relevant amplitude, the shot-noise reference corresponded to the power of the balanced homodyne detector's local oscillator alone. Varying the phase of the local oscillator enabled quantum tomography on the squeezed states. 
The SPM was realized in a second-order nonlinear crystal whose temperature was set to minimal second-harmonic generation. Limitations to the observable squeeze factor were optical loss, phase noise, and potentially inelastic scattering of carrier light at thermally excited photons inside the medium \cite{Callen1952,Shelby1986,Harry2002a,Voss2004,Dong2008,Cesar2009}. Phase noise arises from imperfect stabilization of the length of the squeezing resonator, which is a problem that should get strongly reduced in miniaturized integrated realizations. 
The problem of inelastic scattering of carrier light is generally mitigated by reducing the light power. The design of miniaturized squeezed-light sources using SPM need to take this issue into account. 

\bibliography{library_RS}

\end{document}